\documentclass[aps,pra,twocolumn,amsmath,amssymb,superscriptaddress,reprint,longbibliography]{revtex4-1}
\usepackage[utf8]{inputenc}
\usepackage{amsmath}
\usepackage{amssymb}
\usepackage{mathrsfs}
\usepackage{graphicx}
\usepackage[colorlinks=true,citecolor=blue]{hyperref}
\usepackage{esint}
\usepackage{units}



\def \beq {\begin{equation}}
\def \edq {\end{equation}}
\def \bes {\begin{subequations}}
\def \eds {\end{subequations}}
\def \veps {\varepsilon}
\def \calt {{\cal{T}}}
\def \ket {\rangle}
\def \bra {\langle}

\begin{document}
\title{Phase-Coherent Heat Circulator Based on Multiterminal Josephson Junctions}

\author{Sun-Yong Hwang}
\affiliation{Theoretische Physik, Universität Duisburg-Essen and CENIDE, D-47048 Duisburg, Germany}

\author{Francesco Giazotto}
\affiliation{NEST, Istituto Nanoscienze--CNR and Scuola Normale Superiore, Piazza San Silvestro 12, 56127 Pisa, Italy}

\author{Björn Sothmann}
\affiliation{Theoretische Physik, Universität Duisburg-Essen and CENIDE, D-47048 Duisburg, Germany}

\date{\today}

\begin{abstract}
We theoretically propose a phase-coherent thermal circulator based on ballistic multiterminal Josephson junctions. The breaking of time-reversal symmetry by either a magnetic flux or a superconducting phase bias allows heat to flow preferentially in one direction from one terminal to the next while heat flow in the opposite direction is suppressed. We find that our device can achieve a high circulation efficiency over a wide range of parameters and that its performance is robust with respect to the presence of disorder. We provide estimates for the expected heat currents for realistic samples.
\end{abstract}

\maketitle

\section{Introduction}
The miniaturization of electronics has led to ever more powerful computers but also to an increased production of heat in computer chips. One possible way to deal with such waste heat is to recover it via thermoelectric energy harvesting~\cite{sothmann_thermoelectric_2015}. Alternatively, one may construct logical circuits that operate with heat rather than charge flow~\cite{paolucci_phase-tunable_2018}. The underlying heat flow can be carried by phonons~\cite{li_colloquium:_2012}, photons~\cite{meschke_single-mode_2006,ronzani_tunable_2018}, or electrons~\cite{giazotto_opportunities_2006}.

An ideal playground to investigate phase-coherent electronic heat flows is mesoscopic superconducting circuits~\cite{martinez-perez_coherent_2014,fornieri_towards_2017}. Phase-dependent heat currents in Josephson junctions were predicted theoretically by Maki and Griffin more than half a century ago~\cite{maki_entropy_1965,maki_entropy_1966} but they have been measured only very recently by Giazotto and Martínez-Pérez~\cite{giazotto_josephson_2012}. Phase-coherent heat flow in superconducting circuits allows for the realization of various caloritronic devices such as heat interferometers~\cite{giazotto_phase-controlled_2012,giazotto_josephson_2012,martinez-perez_fully_2013,fornieri_nanoscale_2016,fornieri_0_2017}, thermal diodes~\cite{martinez-perez_efficient_2013,giazotto_thermal_2013,fornieri_normal_2014,martinez-perez_rectification_2015} and transistors~\cite{giazotto_proposal_2014,fornieri_negative_2016}, heat switches~\cite{sothmann_high-efficiency_2017}, thermal memory~\cite{guarcello_josephson_2018}, and nanoscale refrigerators~\cite{solinas_microwave_2016,hofer_autonomous_2016,vischi_coherent_2018}.

Still missing, however, is a phase-coherent heat circulator, which is a multiterminal device that allows heat to flow, say, clockwise from one terminal to the next but blocks thermal transport in the counterclockwise direction. This is analogous to a microwave or radio-frequency circulator in electronics~\cite{hogan_ferromagnetic_1953}. A possible route to the realization of circulators is the use of chiral edge states in quantum Hall systems~\cite{viola_hall_2014,mahoney_-chip_2017} but the required strong magnetic fields are incompatible with superconductivity. In this paper, we suggest multiterminal Josephson junctions subject to a magnetic flux and/or phase bias as an alternative way to realize efficient thermal circulators. Such junctions are also of interest for simulating topologically nontrivial band structures~\cite{meyer_nontrivial_2017,xie_topological_2017}.

\begin{figure}
\centering
	\includegraphics[width=\columnwidth]{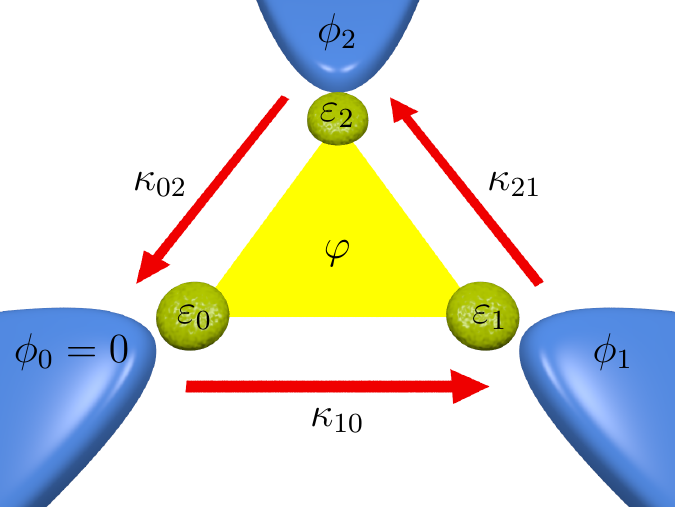}
	\caption{\label{fig:setup} A sketch of the heat circulator based on a minimal three-terminal Josephson junction. A rectified heat circulation with a high efficiency $R$ [cf., Eq.~\eqref{eq:R}] can be achieved with an applied magnetic flux $\varphi$ and two superconducting phases $\phi_1$ and $\phi_2$, i.e., $\kappa_{ij}(\phi_1,\phi_2,\varphi)\gg\kappa_{ji}(\phi_1,\phi_2,\varphi)$. The scattering region is described by a three-site tight-binding model, which is sufficient to capture the essential physics for our proposal.}\label{fig1}
\end{figure}

\section{The Model}
We consider a three-terminal ballistic Josephson junction~\cite{strambini_-squipt_2016} where the junction is described by a minimal model consisting of three
sites that are connected to the three superconducting leads~\cite{meyer_nontrivial_2017} with respective order parameters $\Delta_i=|\Delta_i|e^{i\phi_i}$ (see Fig.~\ref{fig1}). Each site is characterized by an on-site energy $\veps_i$ and hopping matrix elements $t_{ij}$ between the different sites, i.e., from $j$ to $i$. In the absence of a magnetic flux, the hopping matrix elements can be chosen all real. In the presence of a magnetic flux $\varphi$ through the junction, their phases $\gamma_{ij}$, i.e., $t_{ij}=|t_{ij}|e^{i\gamma_{ij}}$, satisfy $\gamma_{10} +\gamma_{21} + \gamma_{02} =2\pi\varphi/\varphi_0 \equiv\alpha$, where $\varphi_0 = h/e$ is the magnetic-flux quantum. By redefining the superconducting phases in a gauge-invariant manner, one can always achieve the symmetric choice $\gamma_{10} = \gamma_{21} = \gamma_{02} = \alpha/3$.
In the following, we consider the simplest case of a symmetric junction with identical on-site energies $\veps_i = \veps$ and identical hopping matrix elements $t_{ij} = t$. The scattering matrix of electrons at the junction is then given by~\cite{meyer_nontrivial_2017} $\hat s_0=\hat U\hat s_D \hat U^\dag$ with $\hat s_D=(1-i\hat D)^{-1}(1+i\hat D)$, where $\hat D=\text{diag}(\veps+2t\cos\frac{\alpha}{3},\veps+2t\cos\frac{\alpha+2\pi}{3},\veps+2t\cos\frac{\alpha-2\pi}{3})$ and
\beq
\hat U=\frac{1}{\sqrt3}
\begin{pmatrix}
1 & 1 & 1\\
1& e^{i2\pi/3} & e^{-i2\pi/3}\\
1 & e^{-i2\pi/3}& e^{i2\pi/3}
\end{pmatrix}.
\edq

In order to obtain the scattering matrix of the whole system at an arbitrary energy $\omega>|\Delta_i|$, we make use of the reflection and
transmission coefficients for Andreev reflection given by $3\times3$ matrices at the normal-superconductor interfaces within the Andreev approximation as~\cite{beenakker_universal_1991,*beenakker_erratum:_1992}
\bes
\begin{align}
\hat r_{eh}^A=\hat r_{he}^A&=-\frac{v}{u}\hat{\mathbf 1}\,,\\
\hat r_{eh}'^{A}=(\hat r_{he}'^{A})^*&=\frac{v}{u}\text{diag}(e^{i\phi_0},e^{i\phi_1},e^{i\phi_2})\,,
\end{align}
\begin{multline}
\hat t_{ee}^{A}=(\hat t_{ee}'^{A})^*=(\hat t_{hh}^{A})^*=\hat t_{hh}'^{A}\\
=\frac{\sqrt{u^2-v^2}}{u}\text{diag}(e^{i\phi_0/2},e^{i\phi_1/2},e^{i\phi_2/2})\,,
\end{multline}
\eds
where the coherence factors are given by
\bes
\begin{align}
u&=\sqrt{\frac{1}{2}\left(1+\frac{\sqrt{\omega^2-\Delta^2}}{\omega}\right)}\,,\\
v&=\sqrt{\frac{1}{2}\left(1-\frac{\sqrt{\omega^2-\Delta^2}}{\omega}\right)}\,.
\end{align}
\eds
Here, one of the superconducting phases can be put equal to zero without loss of generality, i.e., $\phi_0\equiv0$, while the remaining two phases $\phi_1$ and $\phi_2$ can have arbitrary values. We have furthermore assumed an equal gap amplitude $|\Delta_i|=\Delta$ to elucidate our discussion more clearly. 
We then define the matrices
\bes
\begin{align}
\check r&=\left(\begin{array}{cc}
\hat 0 & \hat r_{eh}^A\\
\hat r_{he}^A& \hat 0
\end{array}\right),\qquad
\check r'=\left(\begin{array}{cc}
\hat 0 & \hat r_{eh}'^{A}\\
\hat r_{he}'^{A}& \hat 0
\end{array}\right),\\
\check t&=\left(\begin{array}{cc}
\hat t_{ee}^A& \hat 0\\
\hat 0& \hat t_{hh}^A
\end{array}\right),\qquad
\check t'=\left(\begin{array}{cc}
\hat t_{ee}'^{A}& \hat 0\\
\hat 0& \hat t_{hh}'^{A}
\end{array}\right)\,,
\end{align}
\eds
to obtain the total scattering matrix of the junction as
\beq\label{eq:s}
\check s=\check r+\check t'\check s_{\text N}(\hat{\mathbf 1}-\check r'\check s_{\text N})^{-1}\check t
\edq
where the scattering region is described by the matrix
\beq
\check s_N=
\left(\begin{array}{cc}
\hat s_0 & \hat0\\
\hat0& \hat s_0^*
\end{array}\right)\,.
\edq
From the total scattering matrix in Eq.~\eqref{eq:s}, we obtain the transmission function from terminal $j$ to $i$ as $\calt_{ij}(\omega)=\text{Tr}[\check s_{ij}^{\dag}(\omega)\check s_{ij}(\omega)]=\text{Tr}[\check s_{ij}(\omega)\check s_{ij}^{\dag}(\omega)]$. Due to current conservation, the scattering matrix is unitary, i.e., $\check s^\dag\check s=\check s\check s^\dag=\hat{\mathbf 1}$, and the transmission function satisfies the sum rule $\sum_{i}\calt_{ij}=\sum_{j}\calt_{ij}=2$, where the factor 2 comes from the sum over the electron and hole subspace.
Finally, the thermal conductance matrix is given by its elements ($i,j=0,1,2$)
\beq
\kappa_{ij}=\frac{1}{h}\int d\omega~\omega^2[2\delta_{ij}-\calt_{ij}(\omega)]\zeta(\omega)
\edq
where $\zeta^{-1}(\omega)=4k_\text{B}T^2\cosh^2(\omega/2k_\text{B}T)$. Note that $\kappa_{ij}$ describes the thermal conductance from terminal $j$ to $i$.

Onsager reciprocity dictates that $\kappa_{ij}(\phi_0,\phi_1,\phi_2,\alpha)=\kappa_{ji}(-\phi_0,-\phi_1,-\phi_2,-\alpha)$~\cite{onsager_reciprocal_1931}.
However, one can in general achieve $\kappa_{ij}(\phi_0,\phi_1,\phi_2,\alpha)\ne\kappa_{ji}(\phi_0,\phi_1,\phi_2,\alpha)$, i.e., heat rectification in linear response is possible in our setup by adjusting $\phi_1$, $\phi_2$ and $\alpha$ with $\phi_0\equiv0$.
Based on this observation, we propose a heat circulator with which one can make the heat rotate along the junction preferentially in one direction.
We emphasize that thermal rectification and heat circulation can occur in our device in linear response due to its multiterminal nature. In contrast, conventional two-terminal devices can achieve thermal rectification only in nonlinear transport.

\section{Results and discussion}
In order to quantify the circulator efficiency, we introduce a figure of merit defined by
\beq\label{eq:R}
R=\frac{\kappa_{02}\kappa_{21}\kappa_{10}-\kappa_{01}\kappa_{12}\kappa_{20}}
	{\kappa_{02}\kappa_{21}\kappa_{10}+\kappa_{01}\kappa_{12}\kappa_{20}}\,.
\edq
The first term in the numerator, $\kappa_{02}\kappa_{21}\kappa_{10}$, describes the heat circulation in a counterclockwise sense, i.e., $0\to1\to2\to0$, whereas the second term, $\kappa_{01}\kappa_{12}\kappa_{20}$, quantifies the clockwise rotation $0\to2\to1\to0$. If there is no rectification in thermal conductances, i.e., $\kappa_{ij}=\kappa_{ji}$ for all $i,j$, we have $R=0$, as the numerator in Eq.~\eqref{eq:R} vanishes. In an ideal case with a perfectly counterclockwise heat rotation, the clockwise circulation is completely blocked, i.e., $\kappa_{01}\kappa_{12}\kappa_{20}=0$, while $\kappa_{02}\kappa_{21}\kappa_{10}$ remains finite and hence $R=1$. Analogously, one obtains $R=-1$ for a perfectly clockwise heat circulation.

\begin{figure}
\includegraphics[width=\columnwidth]{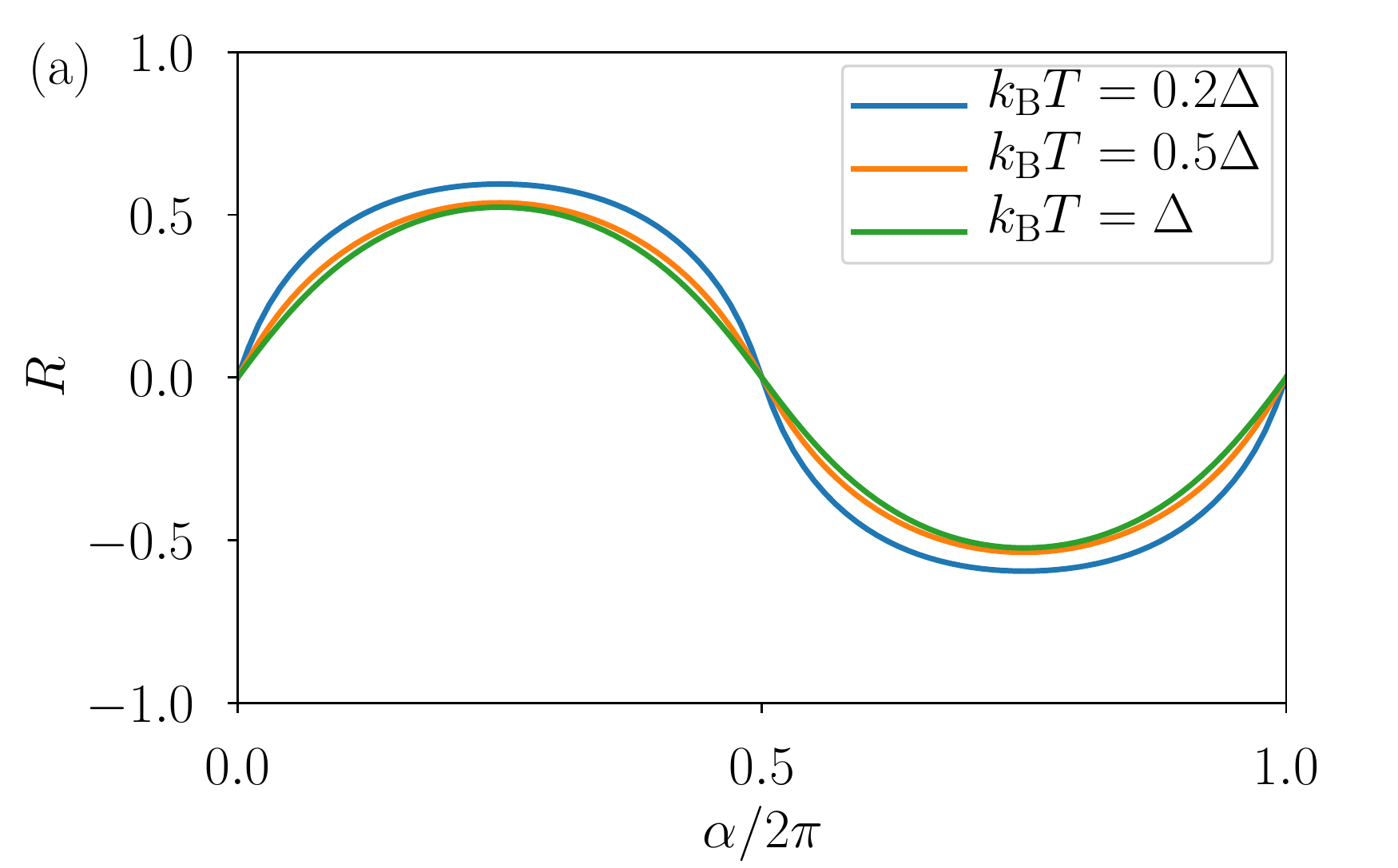}
\includegraphics[width=\columnwidth]{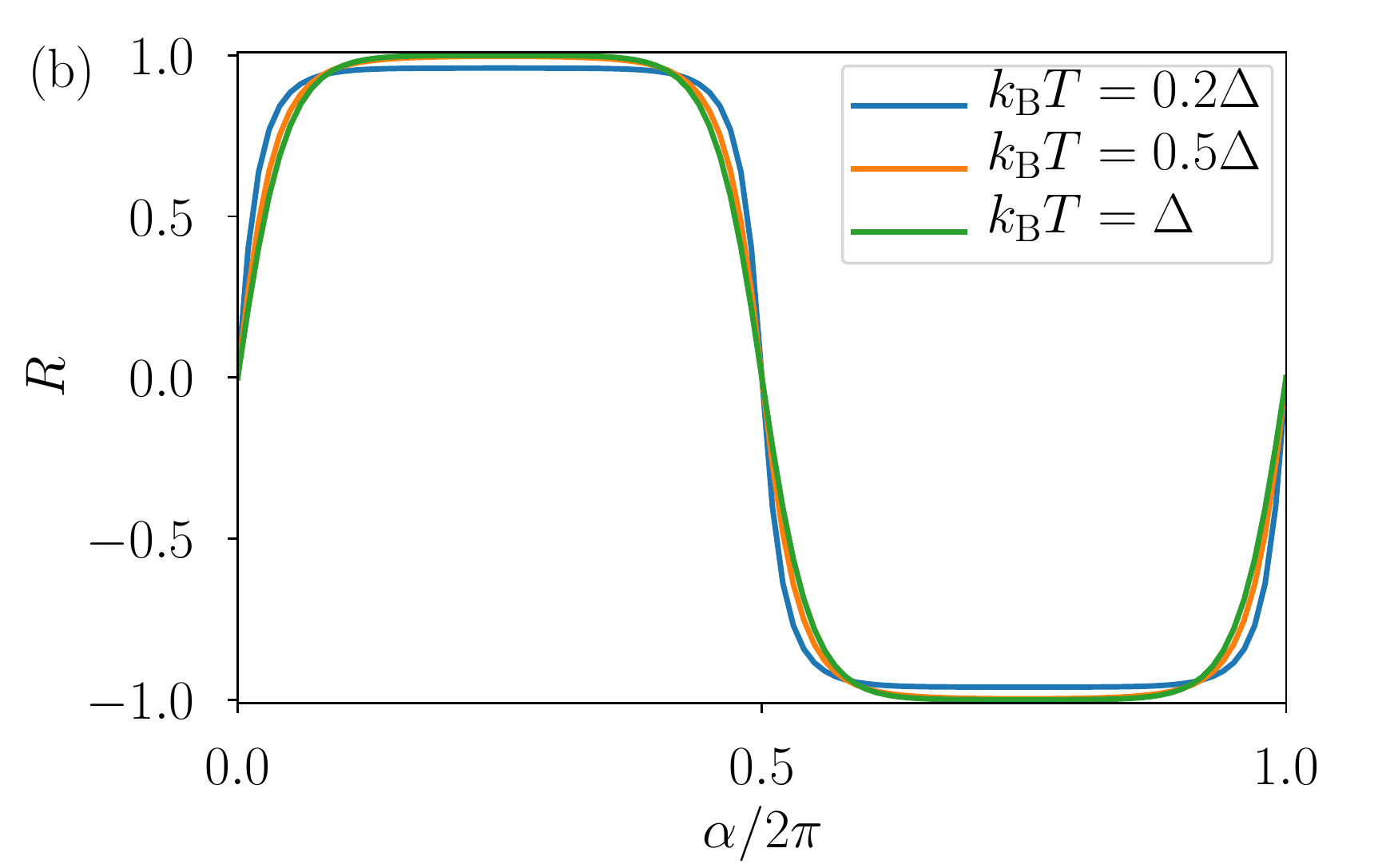}
\caption{$R$ vs $\alpha$ for (a) $t=0.1\Delta$ and (b) $t=\Delta$ with various average temperatures $T$. On-site energies are fixed to the Fermi level $\veps=E_F\equiv0$. No phase bias is applied among the superconductors.}\label{fig2}
\end{figure}
Figure~\ref{fig2}(a) displays $R$ as a function of the magnetic flux $\alpha$ with a relatively weak coupling between the sites, i.e., $t=0.1\Delta$, for several background temperatures $T$. We remark that in linear response the temperature dependence of the order parameter can be neglected with $\Delta=\Delta(T)$. The ratio $T/\Delta$ can be tuned arbitrarily by varying $T$. As $\alpha$ increases, $R$ slowly rises with a positive sign corresponding to the counterclockwise heat circulation and reaches the maximum $R\approx0.6$ at $\alpha=\pi/2$. For $\alpha>\pi/2$, $R$ starts to decrease until $R=0$ at $\alpha=\pi$. For $\pi<\alpha<2\pi$, $R$ becomes negative and the heat rotates clockwise. Note that the efficiency $|R|$ becomes higher with a lower temperature $T$. If we look at the rectification in only one leg of the junction $j\leftrightarrow i$, the maximum efficiency at $\alpha=\pi/2$ corresponds to $\kappa_{ij}/\kappa_{ji}\sim2$ at $k_\text{B}T=0.2\Delta$.

When the coupling strength $t$ gets larger, the efficiency grows more rapidly as $\alpha$ increases and the maximum approaches closer to the ideal case $|R|\sim1$, as shown in Fig.~\ref{fig2}(b) with an example of $t=\Delta$. In addition, the $T$ dependence is reversed at the maximum where a higher $T$ gives rise to a higher efficiency. Indeed, $\kappa_{ij}/\kappa_{ji}$ reaches from $3$ to $22$ as we increase the temperature from $k_\text{B}T=0.2\Delta$ to $k_\text{B}T=\Delta$ [cf., Fig.~\ref{fig4}(a)]. This is a nice property for practical applications, since more quasiparticles are activated when the temperature becomes of the order of the superconducting gap, contributing largely to the thermal conductance. Therefore, our proposal suggests an easy manipulation of the heat rectification, e.g., controlling the direction and the magnitude, with relatively small magnetic fields of the order of millitesla.
This realizes an almost ideal phase-coherent thermal circulator, which so far is still absent in the field of coherent caloritronics.

\begin{figure}
\begin{center}
\includegraphics[width=\columnwidth]{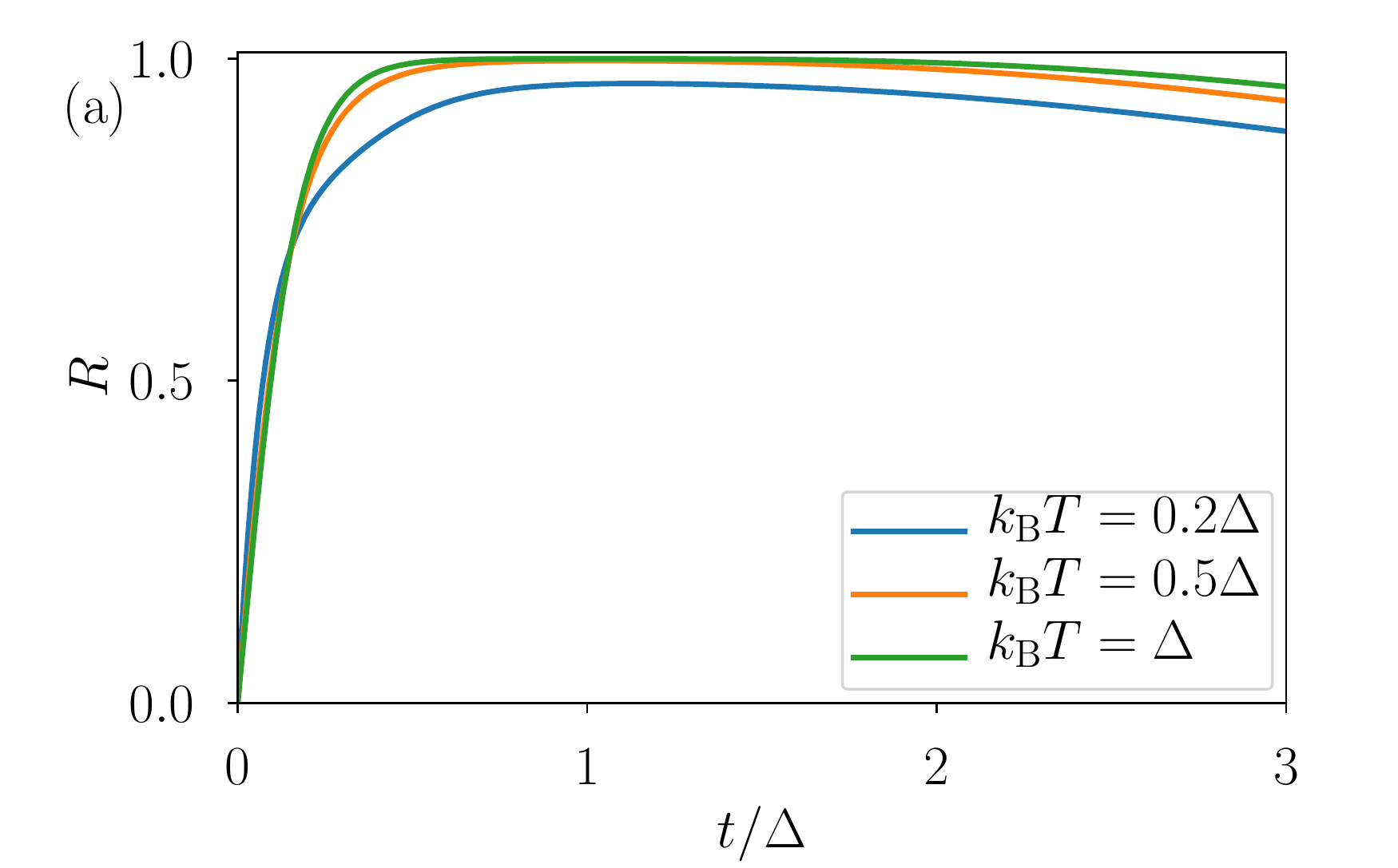}
\includegraphics[width=\columnwidth]{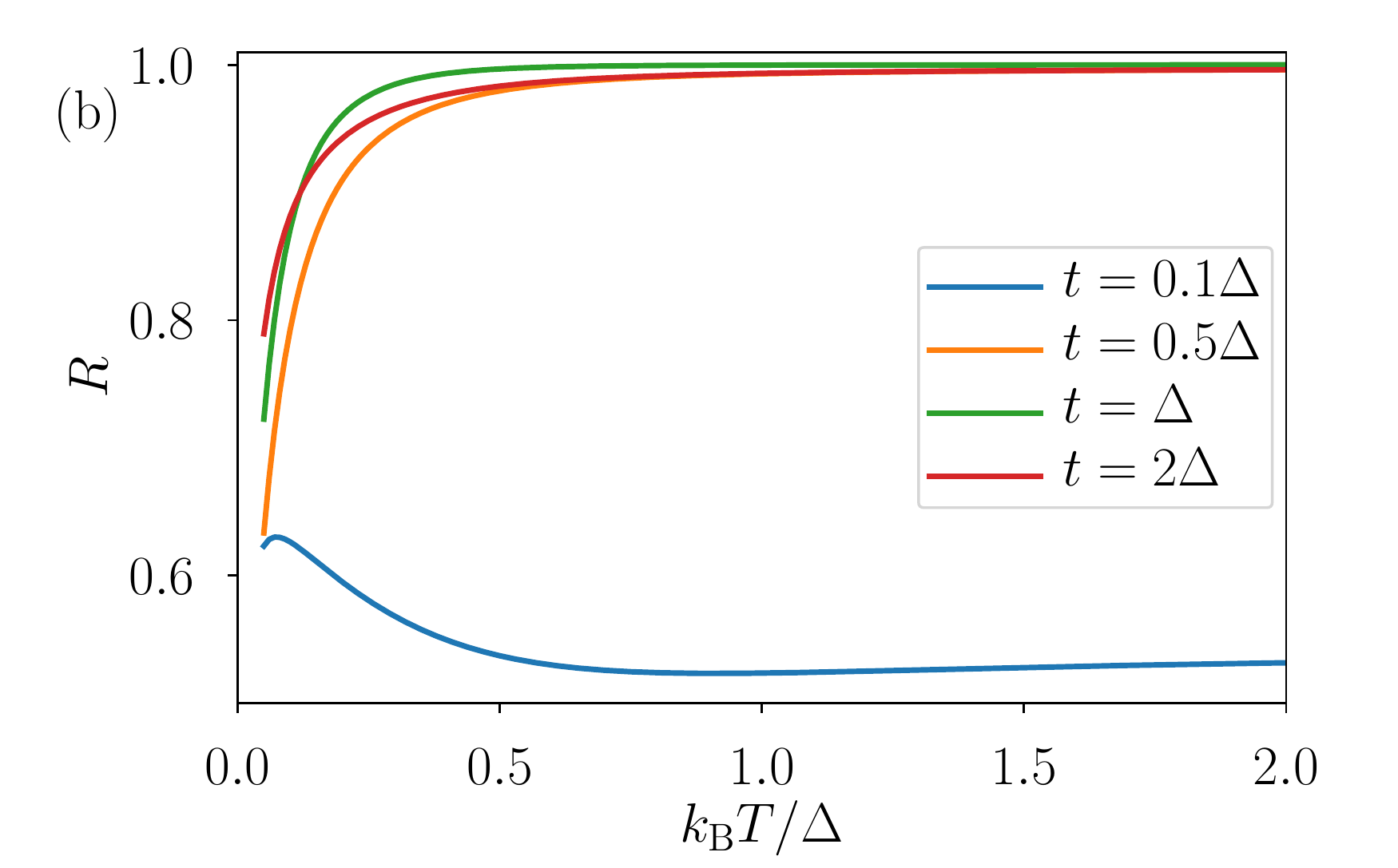}
\caption{$R$ vs (a) $t$ for various $T$ and (b) $T$ for various $t$, at $\alpha=\pi/2$ and $\veps=E_F\equiv0$ with no phase bias.}\label{fig3}
\end{center} 
\end{figure}
In Fig.~\ref{fig3}, we plot the figure of merit $R$ as a function of the coupling strength $t$ and the temperature $T$ at $\alpha=\pi/2$, where $R$ reaches the maximum with a counterclockwise rotation (cf., Fig.~\ref{fig2}). One can note that over a broad range of the parameters, $R$ reaches almost unity as long as $t>0.5\Delta$ [Fig.~\ref{fig3}(a)] and $k_\text{B}T>0.2\Delta$ [Fig.~\ref{fig3}(b)]. Thus, our setup does not require a fine tuning of the control parameters to achieve a high rectification efficiency once these conditions are fulfilled.   

\begin{figure}
\begin{center}
\includegraphics[width=\columnwidth]{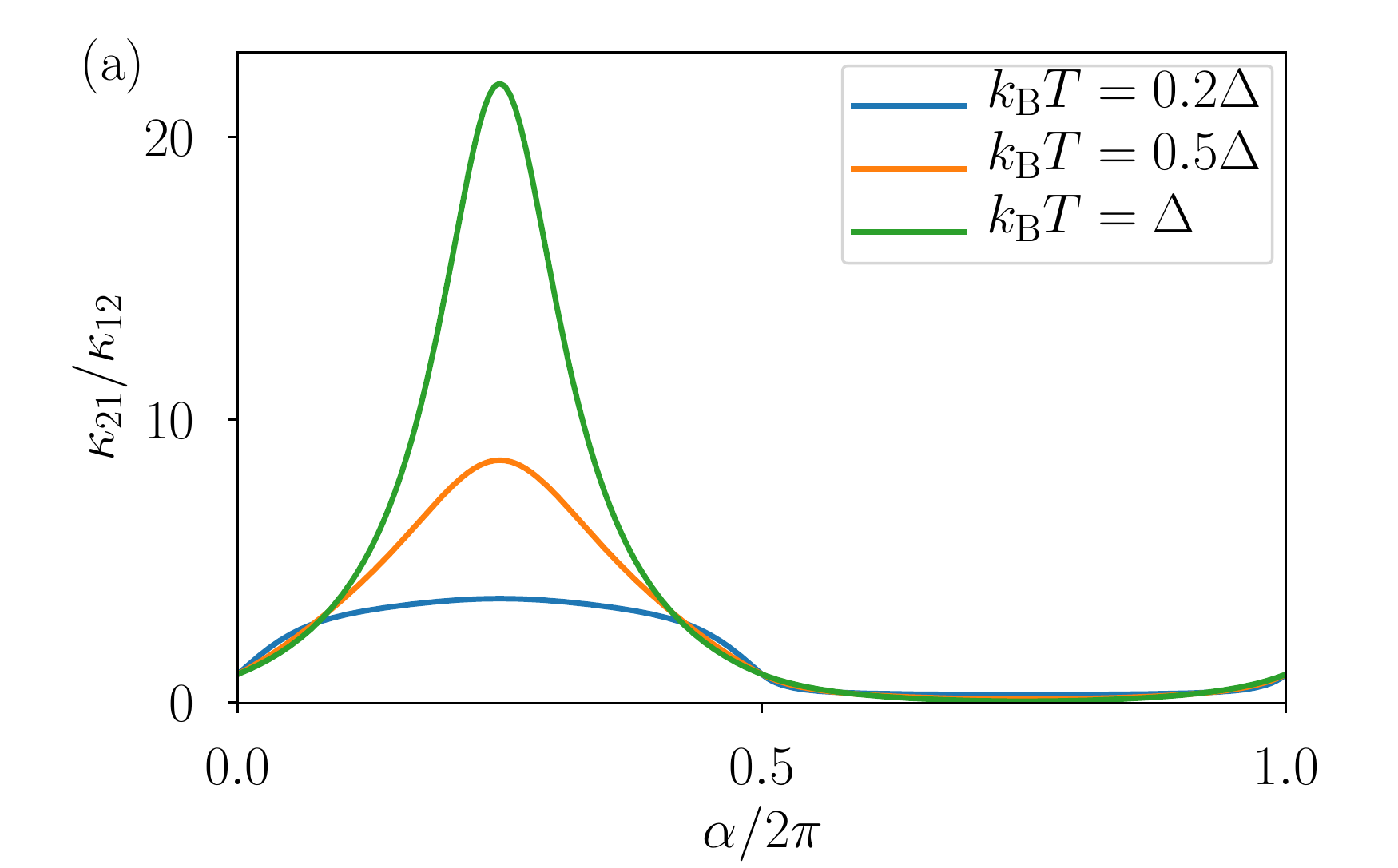}
\includegraphics[width=\columnwidth]{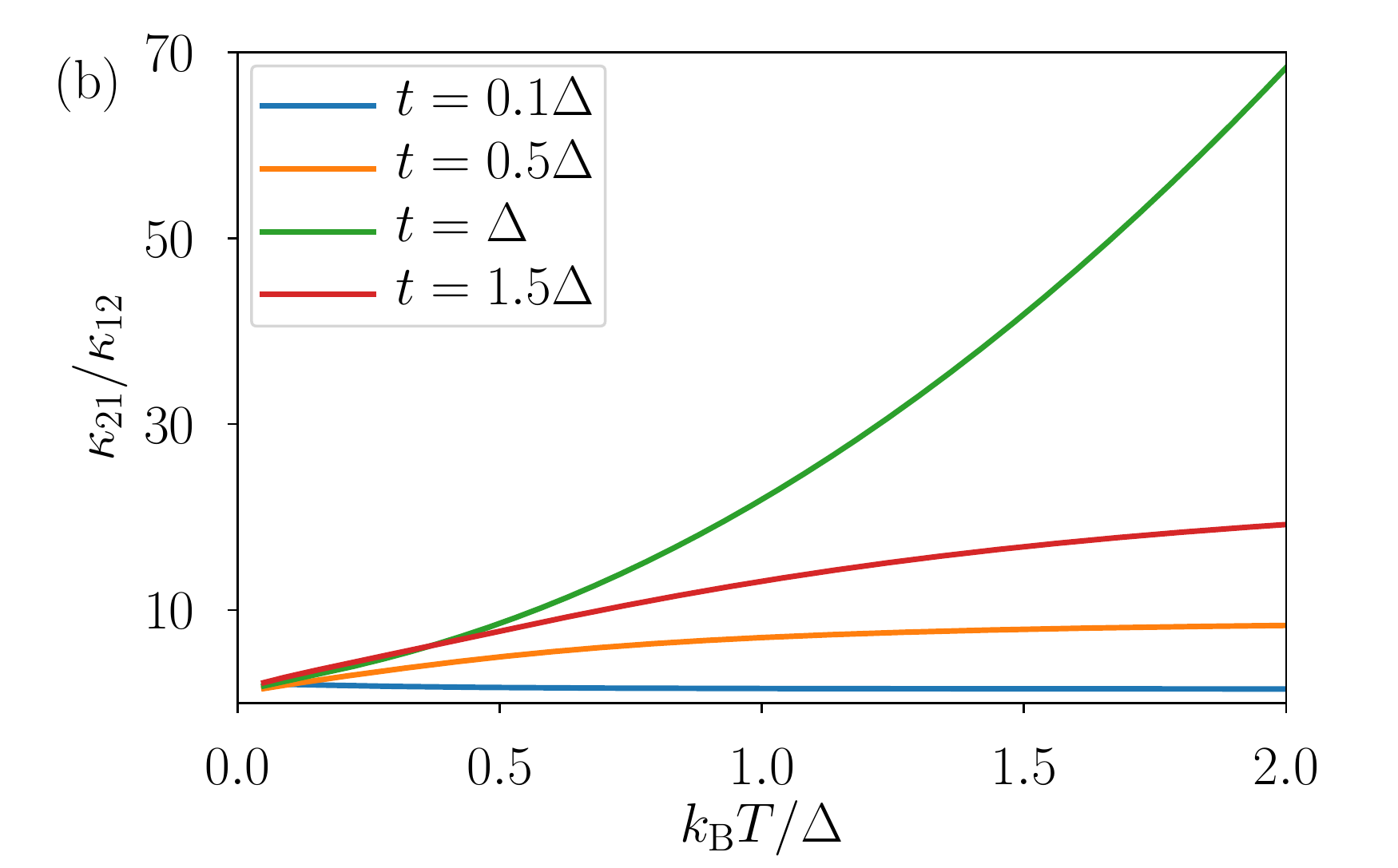}
\caption{$\kappa_{21}/\kappa_{12}$ vs (a) $\alpha$ for various $T$ with $t=\Delta$ and (b) $T$ for various $t$ at $\alpha=\pi/2$. We use $\veps=E_F\equiv0$ and no phase bias is applied among the superconductors.}\label{fig4}
\end{center} 
\end{figure}
In Fig.~\ref{fig4}(a), one-leg rectification $\kappa_{21}/\kappa_{12}$ is shown as a function of $\alpha$ at $t=\Delta$. Indeed, maximum rectification around a factor of $22$ can be achieved with $k_\text{B}T\sim\Delta$ at $\pi/2$, i.e., $22$ times more heat flows preferentially in one direction than in the opposite direction. Remarkably, for this choice of the coupling strengh--of the order of the energy gap, i.e., $t\sim\Delta$--one can obtain an enormous rectification for $k_\text{B}T>\Delta$, e.g., $\kappa_{21}/\kappa_{12}\sim70$ at $k_\text{B}T\sim2\Delta$, as displayed in Fig.~\ref{fig4}(b). However, for $t>\Delta$, the asymmetry $\kappa_{ij}/\kappa_{ji}$ decreases again for any given $T$, e.g., $t=1.5\Delta$ in Fig.~\ref{fig4}(b). We emphasize that $t\sim\Delta$ is actually the optimum coupling strength for enhancing the rectification with our choice $\veps=E_F$. This can be explained by the band dispersion with a width $2t$ in the scattering region, which is commensurate with a superconducting gap of width $2\Delta$, the edge of which possesses a divergent quasiparticle density of states. Therefore, while $\kappa_{ij}/\kappa_{ji}$ saturates as $T$ increases for either $t<\Delta$ or $t>\Delta$, this is not the case for the optimum coupling condition. In a real superconductor, $\Delta$ is not constant but, rather, a decreasing function of $T$ until the superconductivity breaks down at the critical temperature $T_c$, i.e., $\Delta(T_c)=0$. Thus, the relative ratio $k_\text{B}T/\Delta(T)$ can still be much larger than 1 even in a superconducting state, making it possible to observe the huge $\kappa_{ij}/\kappa_{ji}$ as expected from Fig.~\ref{fig4}(b).

\begin{figure}
\centering
\begin{tabular}{cc}
\includegraphics[width=\columnwidth]{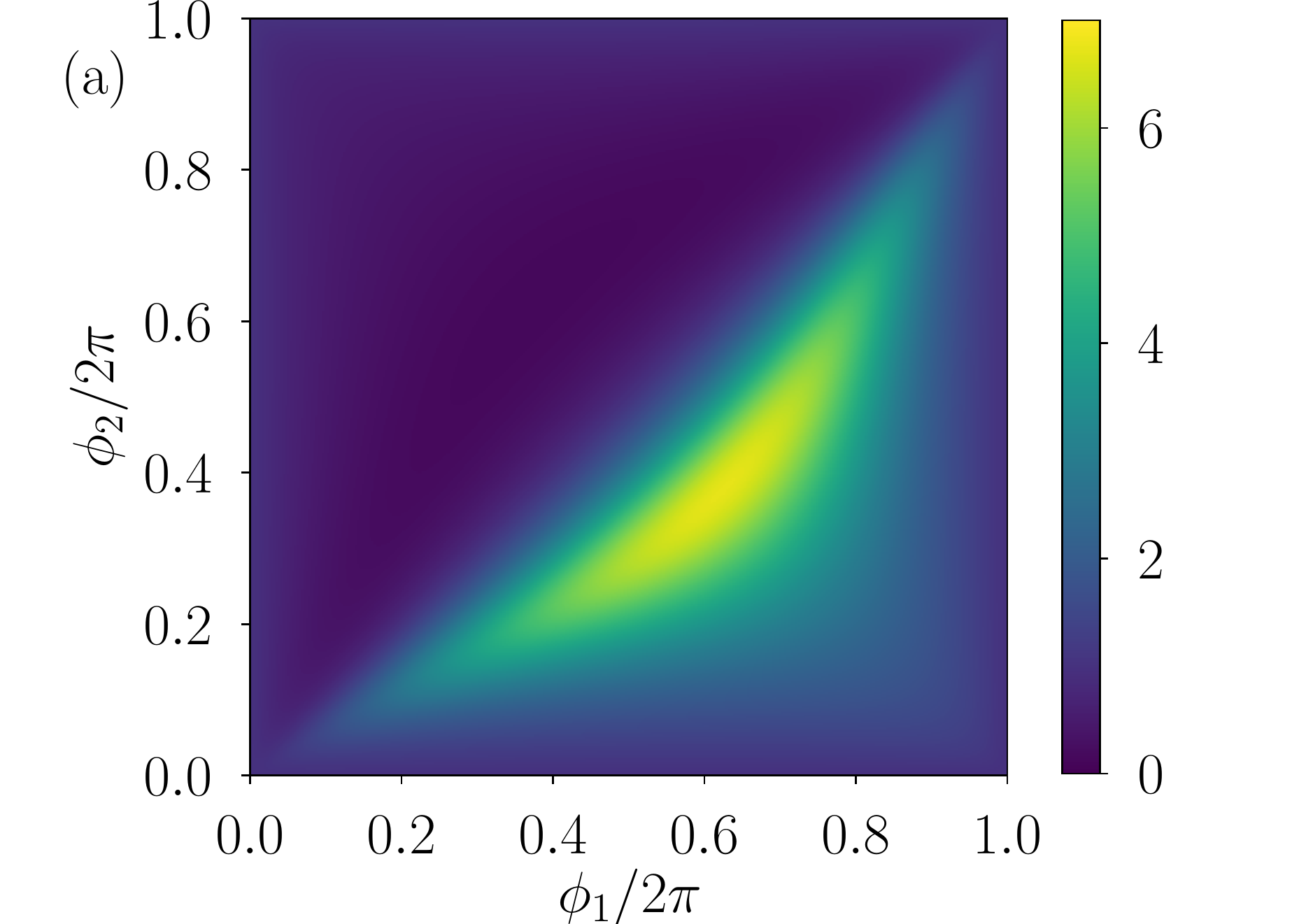}\\
\includegraphics[width=\columnwidth]{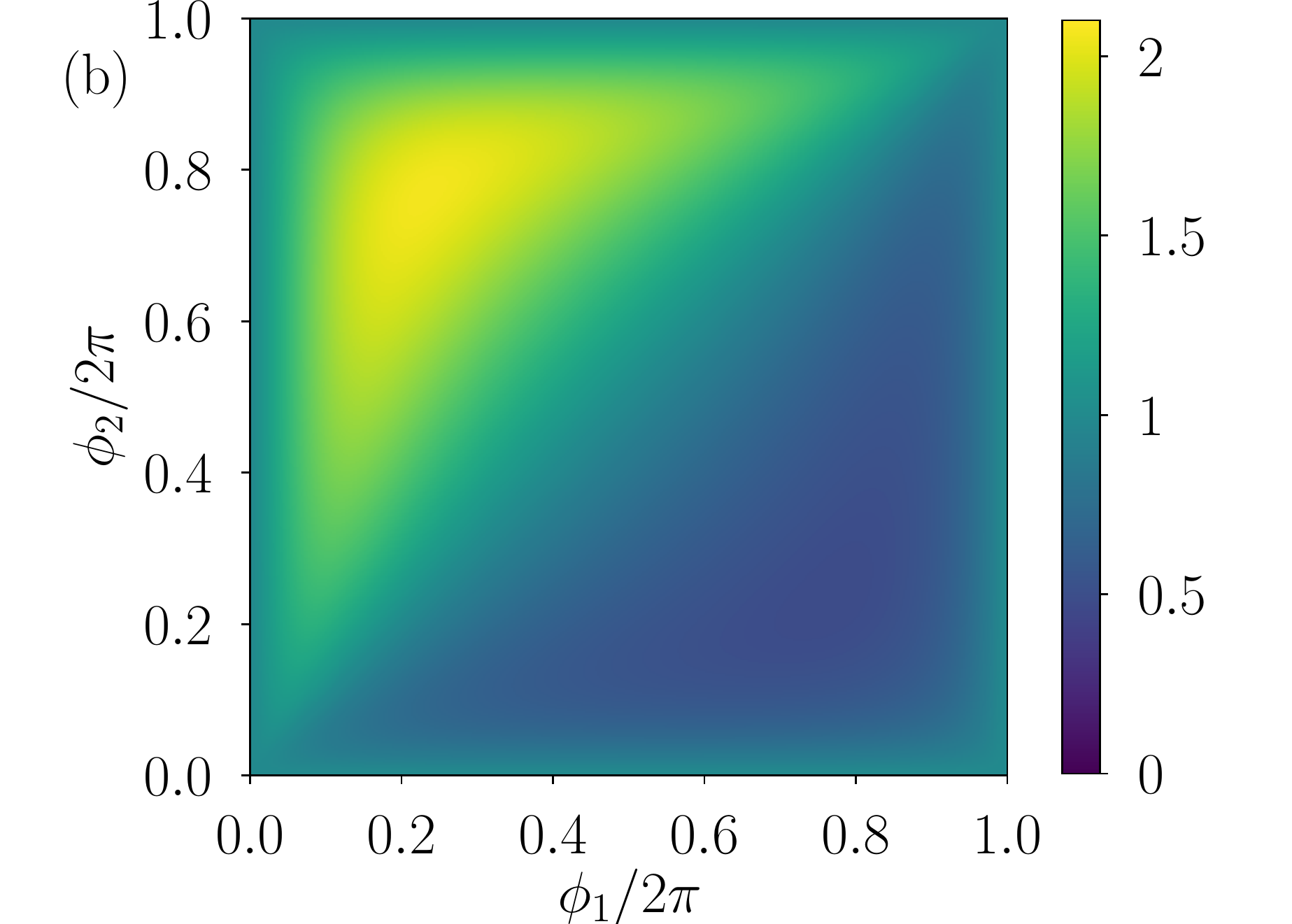}
\end{tabular}
\caption{$\kappa_{21}/\kappa_{12}$ vs $\phi_1$ and $\phi_2$ for (a) $t=0.1\Delta$ and (b) $t=\Delta$, at $k_\text{B}T=0.1\Delta$. The other parameters are $\veps=E_F\equiv0$ and $\alpha=0$.}\label{fig5}
\end{figure}
A phase bias $\phi_1-\phi_2\ne0$ across the junction can also lead to heat rectification. Experimentally, it can be realized in a multiloop structure, where each magnetic flux can be controlled in an independent way.
Figures~\ref{fig5}(a) and \ref{fig5}(b), respectively, display a density plot of $\kappa_{21}/\kappa_{12}$ for $t=0.1\Delta$ and $t=\Delta$ at $k_\text{B}T=0.1\Delta$. For the former, one can obtain $\kappa_{21}/\kappa_{12}\sim6$, while $\kappa_{21}/\kappa_{12}\sim2$ for the latter at an optimum phase bias. In stark contrast to the heat asymmetry controlled by the flux $\varphi$, higher rectification efficiencies can be achieved at lower coupling strengths $t$ and by lowering the temperature $T$.

\begin{figure}
\includegraphics[width=\columnwidth]{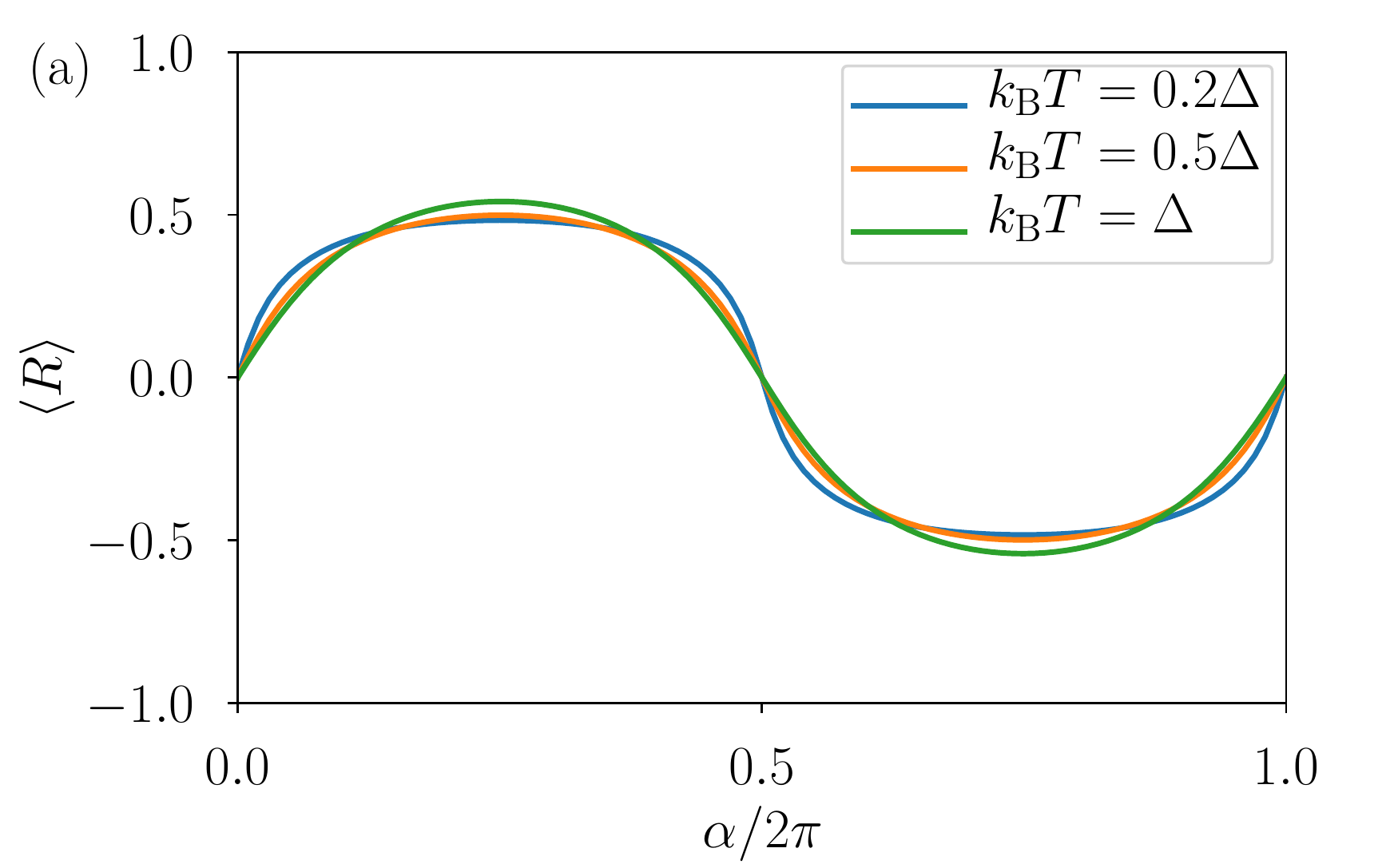}
\includegraphics[width=\columnwidth]{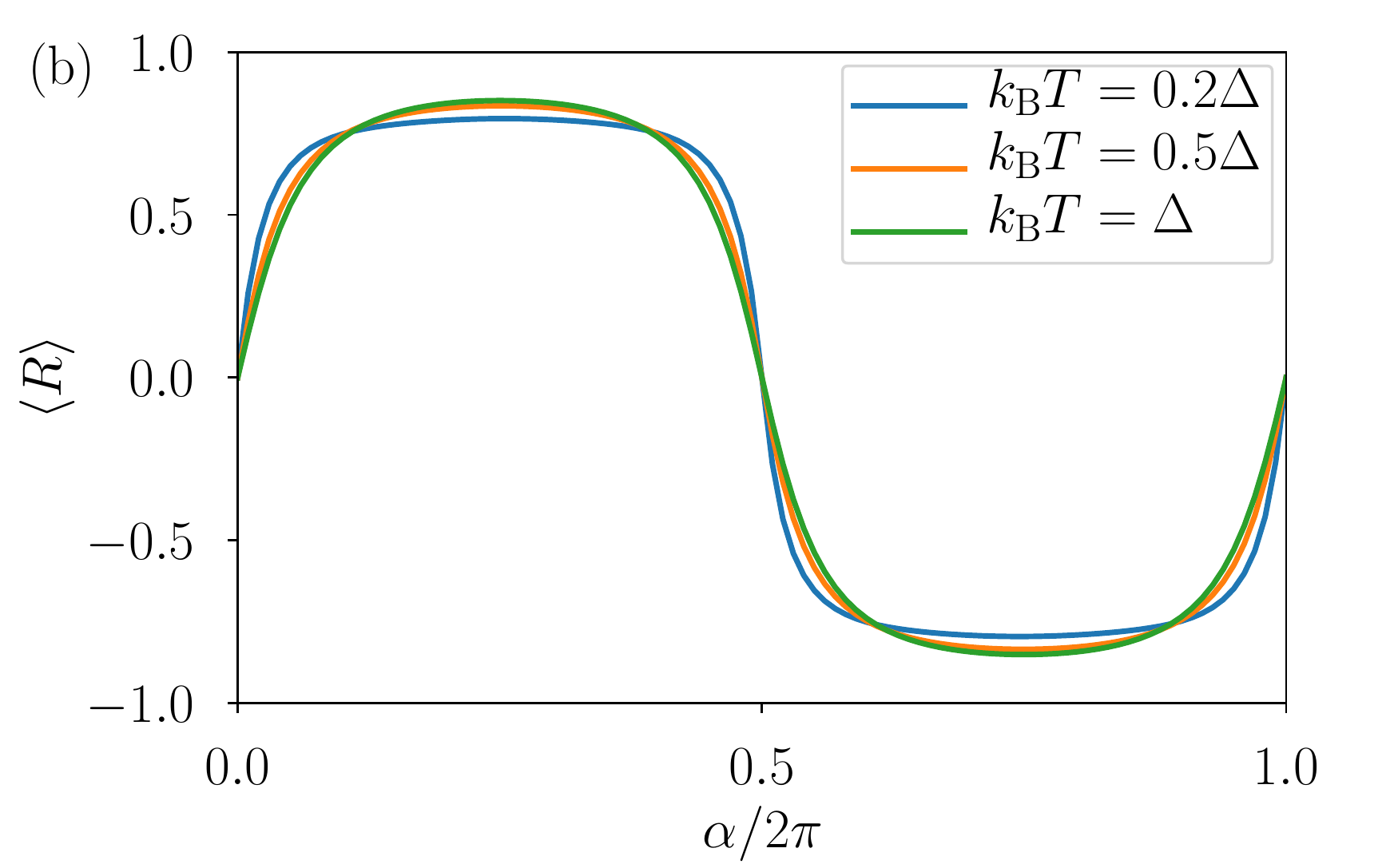}
\caption{$\bra R\ket$ vs $\alpha$ with random parameters averaged over $1000$ samples for (a) $\bra t_{ij}\ket=0.1\Delta$ and $\text{w}(t_{ij})=0.2\Delta$ and (b) $\bra t_{ij}\ket=\Delta$ and $\text{w}(t_{ij})=2\Delta$ with various average temperatures $T$. Average on-site energies are fixed to the Fermi level $\bra\veps_{i}\ket=E_F\equiv0$ with the variation within the full width $\text{w}(\veps_{i})=2\Delta$, i.e., $\bra\veps_{i}\ket-\text{w}(\veps_{i})/2<\veps_i<\bra\veps_{i}\ket+\text{w}(\veps_{i})/2$ and $\veps_i\ne\veps_j$ for all $i,j$.}\label{fig6}
\end{figure}
Thus far we have considered a perfectly symmetric junction with equal on-site energies $\veps_i = \veps$ and hopping matrix elements $t_{ij} = t$. Relaxing this condition to the asymmetric case with arbitrary $\veps_i$ and $t_{ij}$ is straightforward~\cite{meyer_nontrivial_2017}.

Figure~\ref{fig6} shows the effect of a strong random variation of the parameters on the rectification efficiency $\bra R\ket$ averaged over $1000$ random samples, which can be compared to the results of symmetric junction in Fig.~\ref{fig2}. Increasing the sample number leads to effectively the same results. In Fig.~\ref{fig6}(a), the respective coupling strength is a uniform random variable with a mean centered at $\bra t_{ij}\ket=0.1\Delta$ and the full width $\text{w}(t_{ij})=0.2\Delta$, i.e., $\bra t_{ij}\ket-\text{w}(t_{ij})/2<t_{ij}<\bra t_{ij}\ket+\text{w}(t_{ij})/2$, with $t_{ij}\ne t_{k\ell}$ for all $i,j,k,\ell$. On-site energy fluctuations are analogously described by $\bra\veps_i\ket=E_F$ and $\text{w}(\veps_i)=2\Delta$. Likewise, in Fig.~\ref{fig6}(b), we introduce stronger randomness with $\bra t_{ij}\ket=\Delta$ and $\text{w}(t_{ij})=2\Delta$. In both cases, the averaged efficiency $\bra R\ket$ tends to decrease by roughly 15\% compared to the symmetric case due to the strong disorder with the random fluctuation of wide ranges $\text{w}(\veps_i)=2\Delta$ and $\text{w}(t_{ij})=2\bra t_{ij}\ket$. Instead, if one reduces the fluctuations by choosing $\text{w}(\veps_i)=\Delta$ and $\text{w}(t_{ij})=\bra t_{ij}\ket$, the figure of merit $R$ quickly recovers the value of the symmetric junction with a reduced efficiency drop. Hence, the high rectification efficiency of our proposed heat circulator is rather robust with respect to the unintended random variations of the parameters.
Moreover, even if the system were diffusive rather than ballistic, we do not expect drastic changes in the circulator performance since the main underlying mechanism of our proposal is the broken time-reversal symmetry in multiterminal configurations.

Finally, we estimate the expected heat current in the forward direction for realistic superconductor samples with an average temperature $k_\text{B}T=\Delta$ at $\alpha=\pi/2$ and $t=\Delta$ (cf., Fig.~\ref{fig4}). We assume the linear response temperature gradient \unit[100]{mK} to obtain about \unit[65]{fW} for Al with $\Delta\approx\unit[0.2]{meV}$ and \unit[450]{fW} for Nb-based superconductors with $\Delta\approx\unit[1.5]{meV}$.
Schottky-barrier-free semiconducting two-dimensional electron gases (2DEGs) such as InAs~\cite{giazotto_josephson_2004} or In$_{0.75}$Ga$_{0.25}$As~\cite{deon_proximity_2011,amado_electrostatic_2013,fornieri_ballistic_2013} can provide ideal material systems for the realization of the Josephson thermal circulator, as they allow one to easily achieve the ballistic regime with high semiconductor-superconductor interface transparency.

\section{Conclusions}
In summary, we propose a phase-coherent thermal circulator based on ballistic multiterminal Josephson junctions. Its operation relies on the breaking of time-reversal symmetry by a magnetic flux or a superconducting phase bias. We demonstrate that the device achieves a high circulation efficiency over a wide range of parameters. Furthermore, its operation is robust with respect to the presence of disorder.

\begin{acknowledgments}
We acknowledge fruitful discussions with F.~Hassler, J.-H.~Jiang, R.-P.~Riwar, D.~Sánchez and R.~Sánchez, and financial support from the Ministry of Innovation NRW via the ``Programm zur Förderung der Rückkehr des hochqualifizierten Forschungsnachwuchses aus dem Ausland''.
F.G. acknowledges the European Research Council under the European Union’s Seventh Framework Programme (FP7/2007-2013)/ERC Grant Agreement No. 615187-COMANCHE for partial financial support.
\end{acknowledgments}


%

\end{document}